\documentclass[aps,pra,twocolumn,longbibliography,superscriptaddress]{revtex4-2}
   
\usepackage[table]{xcolor}
\usepackage{booktabs}
\usepackage[utf8]{inputenc}
\usepackage[T1]{fontenc}
\usepackage{lineno}

\usepackage[normalem]{ulem}
\usepackage{amsmath}
\usepackage{amsthm}
\usepackage{amssymb}
\usepackage{mathtools} 
\usepackage{bm}
\usepackage{braket}
\usepackage{lineno}
\usepackage{pifont}
\usepackage{bbm}
\usepackage{bbding}
\usepackage{hyperref}

\usepackage{algorithm} 
\usepackage{algpseudocode} 
\usepackage{graphicx}
\usepackage{dcolumn}
\usepackage{bm}
\usepackage{amsmath}
\usepackage{subfigure}
\usepackage{comment}
\usepackage{tikz}
\usepackage{amsfonts}
\usepackage{physics}
\usepackage{amsfonts}

\newcommand{\overbar}[1]{\mkern 1.5mu\overline{\mkern-1.5mu#1\mkern-1.5mu}\mkern 1.5mu}

\begin{document}

\preprint{APS/123-QED}

\title{Efficient and practical quantum compiler towards multi-qubit systems with deep reinforcement learning}

\author{Qiuhao Chen}
\affiliation{
 School of Mathematics and Statistics, Wuhan University, Wuhan, China 
}
\affiliation{
JD Explore Academy, Beijing, China
}

\author{Yuxuan Du}
\thanks{Corresponding authors}
\affiliation{
JD Explore Academy, Beijing, China
}

\author{Qi Zhao}
\affiliation{Joint Center for Quantum Information and Computer Science,
University of Maryland, College Park, Maryland 20742, USA}

\author{Yuling Jiao}
\affiliation{
 School of Mathematics and Statistics, Wuhan University, Wuhan, China 
}

\author{Xiliang Lu}
\affiliation{
 School of Mathematics and Statistics, Wuhan University, Wuhan, China 
} 

\author{Xingyao Wu}
\thanks{Corresponding authors}
\affiliation{
JD Explore Academy, Beijing, China
}

\date{\today}

\begin{abstract}
Efficient quantum compiling tactics greatly enhance the capability of quantum computers to execute complicated quantum algorithms. Due to its fundamental importance, a plethora of quantum compilers has been designed in past years. However, there are several caveats to current protocols, which are low optimality, high inference time, limited scalability, and lack of universality. To compensate for these defects,  here we devise an efficient and practical quantum compiler assisted by advanced deep reinforcement learning (RL) techniques, i.e., data generation, deep Q-learning, and AQ* search. In this way, our protocol is compatible with various quantum machines and can be used to compile multi-qubit operators.  We systematically evaluate the performance of our proposal in compiling quantum operators with both inverse-closed and inverse-free universal basis sets. In the task of single-qubit operator compiling, our proposal outperforms other RL-based quantum compilers in the measure of compiling sequence length and inference time. Meanwhile,  the output solution is near-optimal, guaranteed by the Solovay-Kitaev theorem. Notably, for the inverse-free universal basis set, the achieved sequence length complexity is comparable with the inverse-based setting and dramatically advances previous methods. These empirical results contribute to improving the inverse-free Solovay-Kitaev theorem. In addition, for the first time, we demonstrate how to leverage RL-based quantum compilers to accomplish two-qubit operator compiling. The achieved results open an avenue for integrating RL with quantum compiling to unify efficiency and practicality and thus facilitate the exploration of quantum advantages.\end{abstract}

\maketitle

\section{Introduction.}
The first-generation quantum computers \cite{wang2017high, wang2018toward, arute2019quantum, gong2021quantum, wu2021strong, sun2021realization} have shown their potential across many scientific domains such as quantum machine learning \cite{harrow2009quantum,havlicek2018supervised,2014Lloyd,du2021efficient,huang2021power,huang2021experimental, 2014Rebentrost,schuld2019quantum,Wang2021towards}, quantum information processing \cite{devoret2013superconducting, barends2014superconducting,du2022Effcient,du2021exploring,gur2021sublinear,wu2021robust}, and quantum simulation \cite{lloyd1996universal, berry2015simulating, georgescu2014quantum, o2016scalable, kokail2019self,yuan2019theory,endo2020variational,low2019hamiltonian}. In general, the power of a quantum chip heavily depends on the efficiency of its quantum compiler. That is, an optimal quantum compiler can translate a high-level quantum algorithm into the hardware-level operations (i.e., assembly language) using the fewest number of instructions on a universal basis set to achieve the highest accuracy \cite{nielsen00}. Owing to its crucial role, huge efforts have been dedicated to devising efficient quantum compilers and understanding their capabilities. On the theoretical side, Solovay-Kitaev theorem \cite{Kitaev_1997} and the subsequent results  \cite{quant-ph/0505030,Harrow_2002} evidence that for the inverse-closed universal basis set, within an arbitrary tolerance $\varepsilon$, the optimal quantum compiler can approximate any unitary transformation by a sequence of basic operators whose length scales with  $O(log^c(1/\varepsilon))$ and $c\geq 1$. Due to the reality that some quantum hardware only permits an inverse-free universal basis set, these results may be not applicable. To this end, an important line of research is deriving the optimal sequence length in the inverse-free setting. A recent study  \cite{bouland2021efficient} has proved an inverse-free version of Solovay-Kitaev theorem, which states that the sequence length scales with $O(log^c(1/\varepsilon))$ but has a larger $c=8.62$ compared to the inverse-closed setting. Nevertheless, it is still an open question whether the value of $c$ could be further reduced.

\begin{table*}
\renewcommand\arraystretch{1.1}
\colorbox{lime!10}{
\begin{minipage}{1.0\textwidth}
\caption{\label{tab:table1}\small{\textbf{Comparison of our RL-based quantum compiler with other compilers based on RL or conventional strategies along the basis set, scaling (number of qubits) and the length complexity of the compiled sequence.}}}
\setlength{\tabcolsep}{3.5mm}
\small{
{\begin{tabular}{lllll}
\toprule
\textbf{Compiling Method} & \textbf{Basis Set} & \textbf{Scaling} & \textbf{Output Length Complexity} \\
\midrule
Optimality in theory \cite{quant-ph/0505030} & -- & -- & ${O}\left(C\log ^{{1.000}}(1 / \varepsilon)\right)$ \\
Brute force \cite{Zhang_2020} & Fibonacci anyons & Single-qubit & ${O}\left(2.24\log ^{1.43}(1 / \varepsilon)\right)$ \\
Inverse-closed S-K theorem \cite{Zhang_2020} & Fibonacci anyons & Single-qubit & ${O}\left(\log ^{5.18}(1 / \varepsilon)\right)$ \\ 
Inverse-closed S-K theorem \cite{kliuchnikov2013asymptotically} & -- & Single-qubit & ${O}\left(\log ^{3}(1 / \varepsilon)\right)$ \\ 
Inverse-free S-K theorem \cite{bouland2021efficient} & -- & Single-qubit & ${O}\left(\log ^{8.62}(1 / \varepsilon)\right)$ \\ 
Inverse-free algorithm \cite{zhiyenbayev2018quantum} & Inverse-free diffusive set & Single-qubit & ${O}\left(\log ^{1.585}(1 / \varepsilon)\right)$ \\ 
\textbf{Our RL-based compiler} &  \textbf{Inverse-free diffusive set} & \textbf{Single-qubit} & \boldmath${O}\left(2.683\log ^{{0.9735}}(1 / \varepsilon)\right)$\unboldmath \\
Other RL-based compiler \cite{Zhang_2020} & Fibonacci anyons & Single-qubit & ${O}\left(1.55\log ^{1.6}(1 / \varepsilon)\right)$ \\
\textbf{Our RL-based compiler} &  \textbf{Fibonacci anyons} & \textbf{Single-qubit} & \boldmath${O\left(0.737\log ^{{1.52}}(1 / \varepsilon)\right)}$\unboldmath \\
Other RL-based compiler \cite{moro2021quantum} &  HRC efficient universal set & Single-qubit & ${O}\left(\log ^{{1.25}}(1 / \varepsilon)\right)$ \\
\textbf{Our RL-based compiler} &  \textbf{HRC efficient universal set} & \textbf{Single-qubit} & \boldmath${O}\left(0.89\log ^{{1.077}}(1 / \varepsilon)\right)$\unboldmath \\
\textbf{Our RL-based compiler} &  \textbf{HRC efficient universal set} & \textbf{Two-qubit} & \boldmath${O}\left(1.905\log ^{{1.014}}(1 / \varepsilon)\right)$\unboldmath \\
\bottomrule
\end{tabular}}
}
\end{minipage}}
\end{table*}

Despite the theoretical guarantee, it remains obscure how to design such an optimal quantum compiler, precluded by the exponentially scaled search space with respect to the sequence length and qubit number. Previous literature related to the quantum compiler design can be cast into two main categories. The first category covers the quantum compilers that are deterministic with the theoretical guarantee, while they are not universal and can not be generalized. Concretely, quantum compilers proposed in Refs.~\cite{kliuchnikov2012fast, kliuchnikov2013asymptotically, ross2014optimal, kliuchnikov2015practical, selinger2012efficient} attain the optimal gate sequence length $L$ scaling with ${O\left(log(1/\varepsilon)\right)}$ under the \textit{Clifford}+\textit{T} basis set when the target unitary is specified to be a rotational single-qubit gate along the $z$-axis. Similarly, other quantum compilers \cite{2012, wiebe2013floating, jones2013distillation, PhysRevA.88.042325, Bocharov_2013, Bocharov_2015, Bocharov_2015b} approach the optimal scaling ratio for particular rotational quantum gates by exploiting the additional resources such as ancillary qubits, special states, and classical feedback. 
The second category contains the optimization-based quantum compilers, which have good universal properties but lack a theoretical guarantee.  These compilers can be exploited to compile any target operation by the specified compiling basis sets. For instance, Refs.~\cite{Venturelli_2018,booth2018comparing} investigated the application of temporal planners in the problem of quantum circuits synthesis; Ref.~\cite{Khatri_2019, rakyta2022efficient, peres2022quantum} proposed the quantum-assisted compilers to convert the quantum compiling problem into a complex optimization problem, including discrete parameters (e.g.,  quantum circuit layouts) and continuous parameters (e.g., the angles in rotational quantum gates). However, these optimization-based compilers fail to show any prominent advantage in the inference time compared with those deterministic quantum compilers.

Envisioned by the intrinsic nature of the \textit{sequential decision making} in both quantum compiling and reinforcement learning (RL) \cite{jordan2015machine,10.1145/3448250}, a nascent approach is resonating these two advanced techniques and then developing efficient quantum compilers. The central concept behind this track is reformulating quantum compiling as an RL task, which is finding the shortest path (i.e., minimum sequence length) to reach the location closest to the destination (i.e., target unitary) \cite{Zhang_2020,moro2021quantum}. In this manner, RL-based quantum compilers embrace two favorable merits over conventional strategies, i.e., the compatibility of different quantum systems and an efficient inference process.  However, a common caveat of the current RL-based compilers is that all of them can only be applied to the single-qubit scenario. There are multiple reasons resulting in such a dilemma. First, most RL-based compilers generally learn from scratch without leveraging prior knowledge (i.e., sample distribution), which makes learning extremely difficult on high-dimensional quantum systems. Second, current RL-based compilers adopt a simple RL model, which is inferior to fully making use of experience to find the optimal strategy. Last,  during the inference phase, the employed search methods are insufficient to balance efficiency and accuracy, which is substantial for multi-qubit scenarios. With this regard, it is natural to ask: \textit{how to compensate for the above deficiencies and further boost RL-based quantum compilers?} 

\begin{figure}[ht]
\centering
\includegraphics[width=8.7cm]{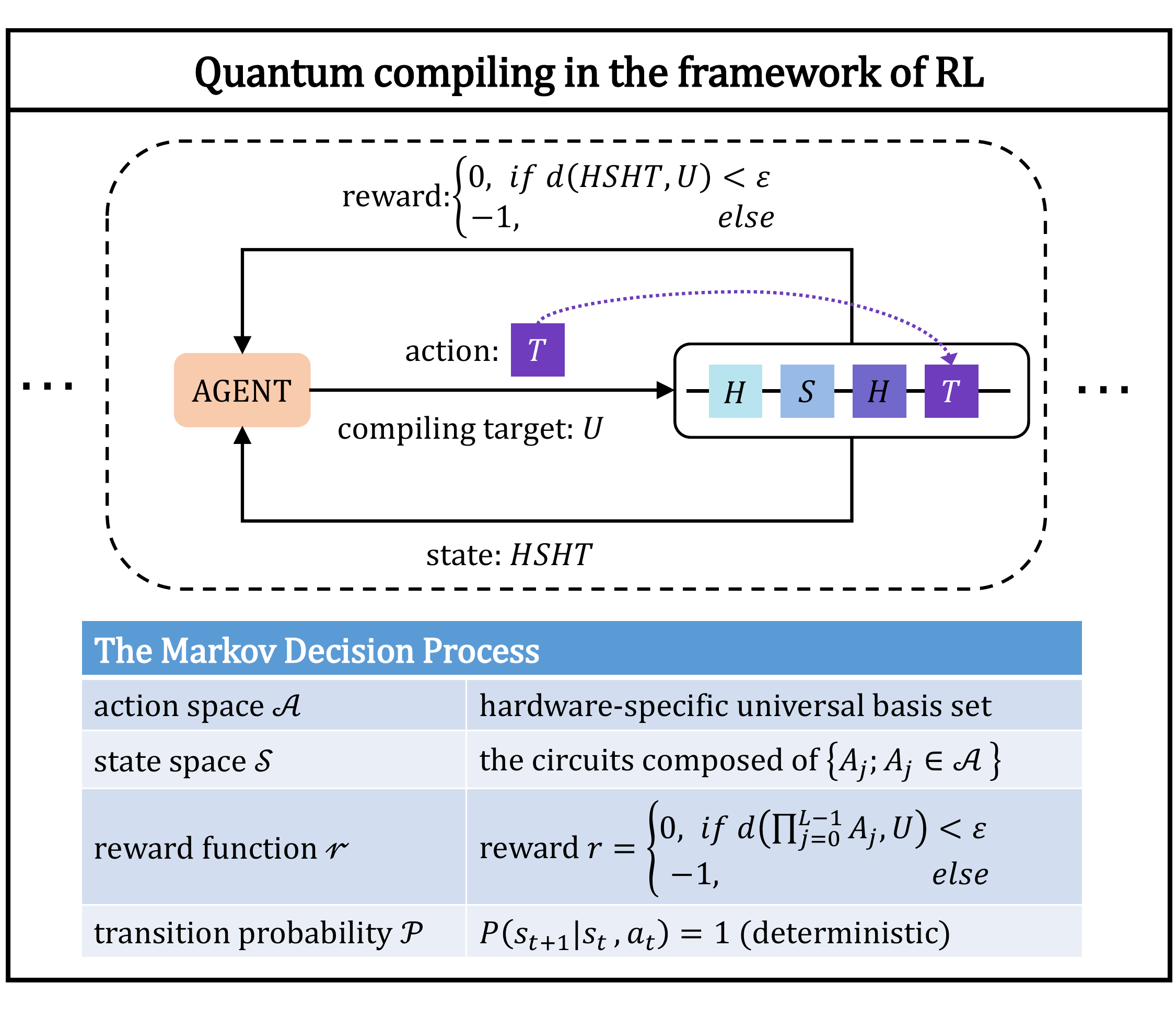} \caption{\small{\textbf{Reformulation of quantum compiling into a RL problem modeled by Markov Decision Process (MDP).} During an interaction between the agent and environment, the agent selects a single gate $T$ from the universal basis set $\mathcal{A}$ to approach the compiling target $U$; the constructed unitary feeds the agent with a new state $HSHT$ as the consequence of the selected gate $T$ and a reward, $0$ or $-1$, measuring the quality of the selected gate sequence $\{H, S, H, T\}$. The MDP ends with a non-negative reward signal $0$, ensuring that the compilation is accomplished. The compiled unitary is formulated by the sequence of the selected gates during this MDP.}
}
\label{fig:reformulation}
\end{figure}

To enhance the power of RL-based quantum compilers, here we devise a deep reinforcement learning-based quantum compiling algorithm with AQ* search. Two key technical components of our proposal are the deep Q-network \cite{mnih2013playing, bellman1966dynamic, puterman1978modified} and the AQ* search strategy \cite{agostinelli2021a, hart1968formal, bonet2001planning}. Intuitively, the former ensures that in the training stage, the compiling experience can be satisfactorily leveraged and ensures that the learned policy well approximates the optimal policy with the theoretical guarantee, which is crucial in compiling multi-qubit gates;  the latter warrants that in the inference stage, the decomposition of the target unitary is accurate and can be completed in an ignorable time. Besides the efficiency, our proposal inherits the compatibility with other RL-based quantum compilers. We conduct systematic numerical simulations to exhibit the superiority of our proposal on both single-qubit and two-qubit operator compiling. To our best knowledge, this is the first time of applying an RL-based quantum compiler to accomplish multi-qubit compiling without any additional restrictions. Specifically, in the task of single-qubit operator compiling,  our algorithm can generate logic quantum operators within a tolerance of $0.99999$ average fidelity under the \textit{inverse-closed} universal basis sets, i.e.,the \textit{Clifford}+\textit{T} universal basis set, Fibonacci anyons basis, and the HRC efficient universal basis set \cite{Harrow_2002},  $0.999$ average fidelity under the \textit{inverse-free} universal basis set \cite{zhiyenbayev2018quantum}, and $0.9996$ average fidelity under the two-qubit universal basis set. For clarity, the comparison among different quantum compilers with respect to the achieved length complexities is summarized in Table \ref{tab:table1}. These results indicate that the solution output by our proposal is \textit{near optimal},  approaching to the lower bound proved in Solovay-Kitaev theorem.

\section{Preliminary}

Before moving on to present our proposal, we first recap  quantum compiling and its reformulation in the language of RL.  Suppose that the target unitary is $U$ and a discrete universal basis set is $\mathcal{A}$.  Mathematically, $\mathcal{A}$ is \textit{inverse-closed} if for every element in this gate set, its exact inverse is also contained; otherwise,  $\mathcal{A}$ is \textit{inverse-free}. 
The purpose of  quantum compiling is to find a minimum sequence of basis $\{A_0, A_1, \dots, A_{L-1}\} \in \mathcal{A}$ such that the distance  between $U$ and $\prod_{j=0}^{L-1} A_{j}$ is bounded within a pre-defined error $\varepsilon$, i.e., 
\begin{eqnarray}\label{eqn:QCL-def}
	d\left(\prod_{j=0}^{L-1} A_{j},U\right)<\varepsilon.
\end{eqnarray}
Throughout the whole study, the distance $d(X, Y)=||X-Y||$ refers to F-norm and $\mathcal{A}$ specifies the universal basis set shown in Table \ref{tab:table1}. Note that our proposal can be easily extended to other universal basis sets and distance measures, e.g., quaternion distance \cite{huynh2009metrics}, diamond norm \cite{aharonov1998quantum}, Hilbert-Schmidt distance \cite{patel2021robust}.

\begin{figure*} 
\centering
\includegraphics[width=0.95\textwidth]{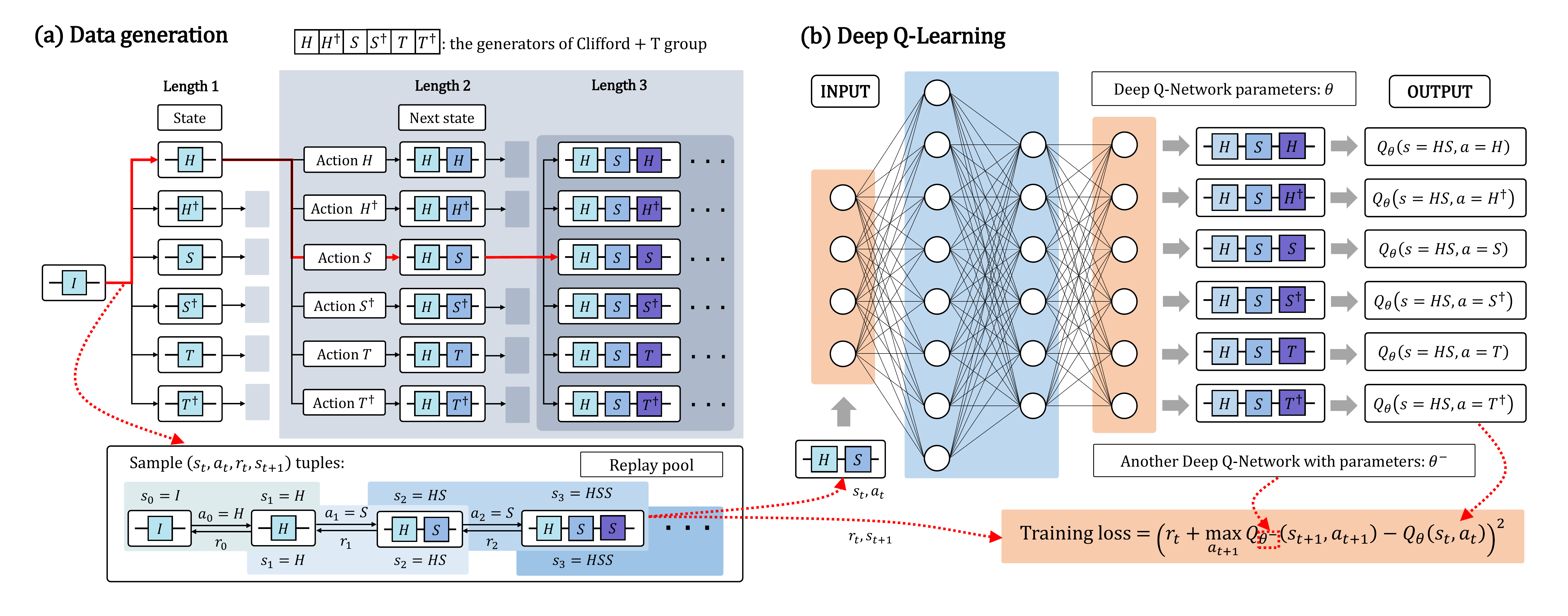}
\caption{\small{\textbf{The process of data generation and deep Q-learning.}
(a) The process of data generation for single-qubit cases under the \textit{Clifford}+\textit{T} universal basis set. The training experience is composed of the sampled trajectories across the whole unitary space, e.g., the trajectory (highlighted by the red line) is randomly sampled from the whole unitary space expanded by a universal basis set. The selected trajectories are put into the replay pool to decouple the correlation between the sample tuples. Namely, a trajectory is decomposed into sequences of sample tuples, and then fed into learning process, e.g., taking action $a_t=S$ from state $s_t=HS$ leads to the next state $s_{t+1}=HSS$ with reward $-1$.   
(b) The process of deep Q-learning amounts to minimizing the training loss at the bottom right plot. The blue (orange) box are hidden layers (input and output layers) of deep Q-networks (DQNs). A key feature of DQNs is that their output layers describe the action-value functions. The input of the DQN is a vector transformed from the unitary matrix corresponding to the state unitary, and the output is a vector corresponding to the value function of all next state unitaries.}
}
\label{fig:data_generation+dqn}
\end{figure*}

The quantum compiling problem in Eq.~\eqref{eqn:QCL-def} can be reformulate into a \textit{value-based RL problem}, which in turn can be addressed by the deep Q-network \cite{mnih2013playing} (detailed in the subsequent context). As shown in Fig.~\ref{fig:reformulation}, the agent proceeds compilation by adding gates sequentially rather than searching a complete gate sequences directly. To avoid the sparse reward issue, our proposal adopts the distance measure $d(U\prod_{j=0}^{L-1}A_j^{\dagger},I)$ instead of $d(\prod_{j=0}^{L-1}A_j^{\dagger},U)$, where $U$ is the  target unitary and $I$ is the identity operator. This tactic is first proposed by \cite{Zhang_2020}.  Using the language of value-based RL,  quantum compiling is equivalent to find the action-value function $Q(s, a)$ obeying the following \textit{Bellman optimality equation},  
\begin{eqnarray}
Q^{*}(s, a)=\mathbb{E}_{s^{\prime} \sim P(\cdot \mid s, a)}\left[r+\gamma \max _{a^{\prime}} Q^{*}\left(s^{\prime}, a^{\prime}\right) \middle| s, a\right],
\label{eq:bellman_optimal}
\end{eqnarray}
where the physical meaning of $s$, $a$, $P$, $r$, and $\gamma$ is state (i.e., $U\prod_{j=0}^{L-1} A_j^{\dagger}$), action (i.e., the adopted quantum gate $A_j\in \mathcal{A}$), transition probability distribution, reward, and the discount factor, respectively. The notation $s^{\prime}$ refers to the next state of the composed unitary obeying the distribution $P(\cdot \mid s, a)$. The optimal action-value function $Q^{*}(s, a)$ is defined as the maximum expected cumulative reward achievable by taking action $a$ when some states $\{s\}$ are observed.  In the task of quantum compiling, {$Q^{*}(s=U^{} \prod_{j=1}^{L}A_j^{\dagger}, a=A_{L+1})$ refers to the negative shortest distance (the minimum sequence length) between the identity state $I$ and the next state $s^{\prime}=U^{} \prod_{j=1}^{L+1}A_j^{\dagger}$, which is obtained by interacting the state $s$ with the action gate $a$. An intuition is shown in Fig.~\ref{fig:data_generation+dqn}(a)  and more details are deferred to Appendix \ref{appendix:QC_in_RL}.  Using the Bellman equation to attain the optimal policy, the connection of the action-value function between the $t$-th iteration and the ($t-1$)-th iteration can be established by the Bellman operator,  i.e.,
\begin{eqnarray}\label{eqn:bell_eqn_update}
	Q^{(t)}(s,a)=\mathbb{E}[r+\gamma \max_{a^{\prime}}Q^{(t-1)}\left(s^{\prime}, a^{\prime}\right) \mid s, a].
\end{eqnarray}
It has been proved that such action-value function converges to the optimal action-value function, $Q^{(t)} \rightarrow Q^{*}$ as $t \rightarrow \infty$  \cite{sutton2018reinforcement}. Compared with prior RL-based strategies, this reformulation can not only alleviate the computational bottleneck but also enables the theoretical guarantee for convergence. Refer to Appendix \ref{append:comp-value-based} for details.

\section{RL-enhanced quantum compiler}  
We now elucidate how our proposal accomplishes the quantum compiling task described in Eq.~\eqref{eqn:bell_eqn_update}. Our proposal consists of three components. At the initialization stage, our protocol  builds an efficient sample distribution that approximates a uniform distribution over each state-action pair. In the training procedure, our protocol employs the   \textit{deep Q-network} (DQN) \cite{mnih2013playing, agostinelli2021a} to minimize the cost function in Eq.~\eqref{eqn:bell_eqn_update}, which efficiently approximates the optimal action-value function $Q^{*}(s, a)$ taking advantage of the generated sample experience. At the inference stage, our protocol exploits AQ* search guided by the trained DQN to seek the optimal gate sequence for a given quantum operator.

The optimization of DQN is as follows. Recall that in the context of quantum compiling, the state space is exponentially scaled with the sequence length and number of qubits. As such, the updating rule in Eq.~(\ref{eqn:bell_eqn_update}) is impractical for multi-qubit systems. To this end,  an alternative is employing a DQN \cite{mnih2013playing, agostinelli2021a} $Q(s, a; \theta^{(t)})$ as shown in Fig.~\ref{fig:data_generation+dqn}(b) to estimate the action-value function $Q^{(t)}(s, a)$ in Eq.~(\ref{eqn:bell_eqn_update}). This approximation is achieved by tuning  trainable parameters $\theta^{(t)}$ to minimize a sequence of loss function $\{L_t(\theta^{(t)});t=0, 1, 2, \cdots \}$    with 
\begin{eqnarray}
L_{t}\left(\theta^{(t)}\right)=\mathbb{E}_{s, a \sim \phi(\cdot)}\left[\left(y^{(t)}-Q\left(s, a ; \theta^{(t)}\right)\right)^{2}\right],
\label{eqn:bellman_loss}
\end{eqnarray}
where $y^{(t)}\!=\!\mathbb{E}_{s^{\prime}\! \sim\! P(\cdot \mid s,a)}\left[r\!+\!\gamma \!\max _{a^{\prime}}\! Q\left(s^{\prime},\! a^{\prime}\! ; \theta^{(t-1)}\right)\! \mid\! s,\! a\right]$ is the target for iteration $t$ and $\phi(s, a)$, a.k.a., \textit{behavior distribution},  can be any probability distribution over states and actions \cite{sutton2018reinforcement, mnih2013playing}. To attain a good learning performance, here we construct the empirical behavior distribution over the $(s, a)$ sample pairs by randomly and sequentially taking actions from the goal state $I$. An intuition of generating trajectories is shown in Fig.~\ref{fig:data_generation+dqn}(a). Besides, the implementation of DQN  allows a stable learning performance. Specifically, we start with a DQN with randomly initialized parameters, and the training experience below a predefined gate sequence length  $d$ is collected to optimize the DQN by minimizing the cost function in Eq.~\eqref{eqn:bellman_loss}. The optimization of DQN with gate sequence length $d$ continuously proceeds until the training loss in Eq.~\eqref{eqn:bellman_loss} is below a pre-fixed threshold $\delta$. Subsequently, the training experience below a predefined gate sequence length $d+1$ is fed into DQN to minimize the cost function in Eq.~\eqref{eqn:bellman_loss}. Similarly, the optimization of DQN continuously proceeds until the training loss reaches a threshold $\delta$. The learning process stops when $d$ reaches  a predefined threshold.

We remark that both behavior distribution $\phi(s, a)$ and the adopted DQN contribute to the superior performance of our protocol. Specifically, $\phi(s, a)$  determines the efficiency of the learning process. Specifically, $\phi(s, a)$ ensures that every generated trajectory is an experience of a successful compilation, which avoids the dilemma of sparse reward \cite{riedmiller2018learning} and allows the production of high-quality sample information. In addition, DQN  fully exploits the sample trajectories, which in turn efficiently approximates the optimal strategy and ensures the achievement of multi-qubit compilation. The process of successively increasing $d$ in optimization ensures learning stability. 

\begin{figure*}
\centering
\includegraphics[width=0.99\textwidth]{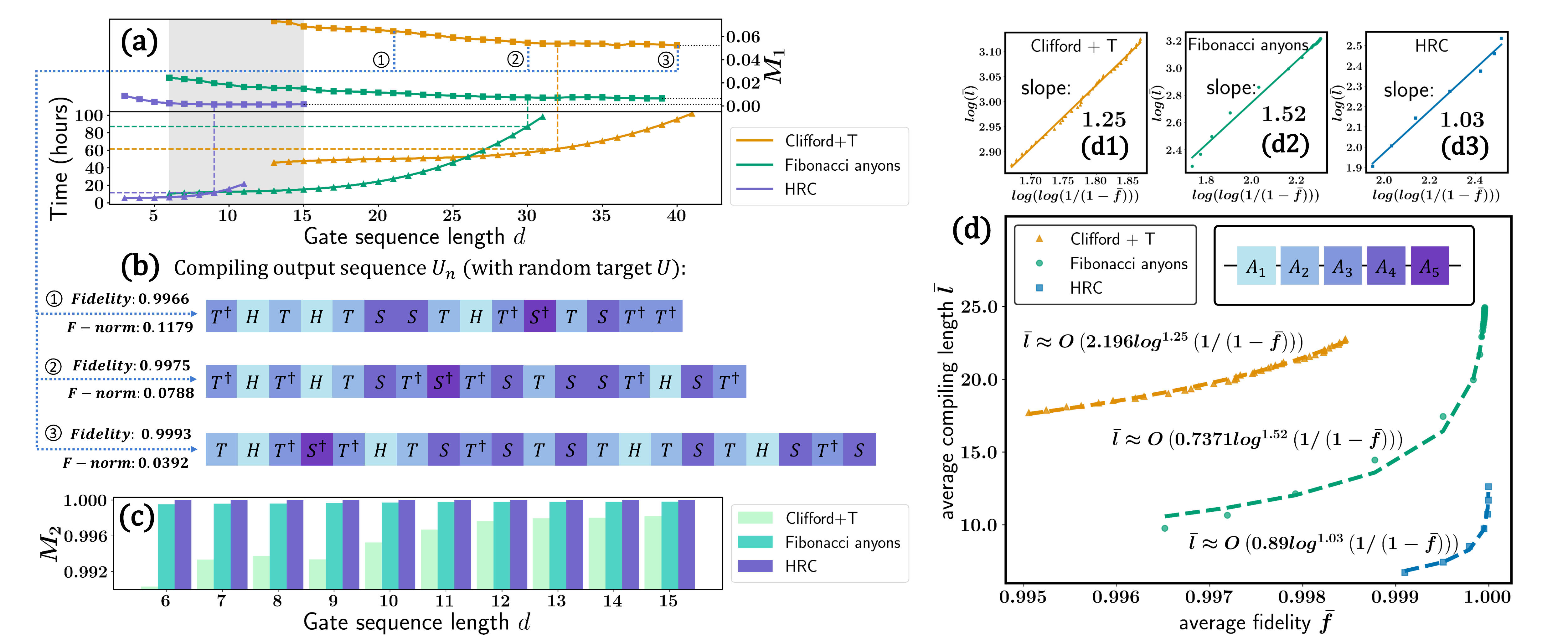} 
\caption{\label{fig:ex_1} 
\small{\textbf{Comparison between different inverse-closed universal basis sets.}  (a) Comparison of accuracy among different universal basis sets. The accuracy is measured by the F-norm metric $M_1$. Every square point is obtained by averaging the distance $M_1(U_n,U)$ of $10^3$ samples, where $U_n$ is generated by AQ* search guided by the trained DQN. Every triangle point denotes the time consumption during the training phase for the DQN expanded by different basis sets with the gate sequence length $d$. (b) The demonstration of the training process indexed by the gate sequence length   $d$ using \textit{Clifford}+\textit{T} universal basis set. (c) The compiling accuracy of the shadow region in (a) using the fidelity metric $M_2$. (d) The top three subplots examine the linearity between $\log\left(\log\left(1/(1-\bar{f}) \right)\right)$ and $\log\left(\bar{l}\right)$ under three basis sets. The lower plot depicts the sequence length complexity versus the average fidelity $\bar{f}$ under different inverse-closed universal basis sets in the single-qubit scenario. 
 }}
\end{figure*} 

\begin{figure}
\centering
\includegraphics[width=0.49\textwidth]{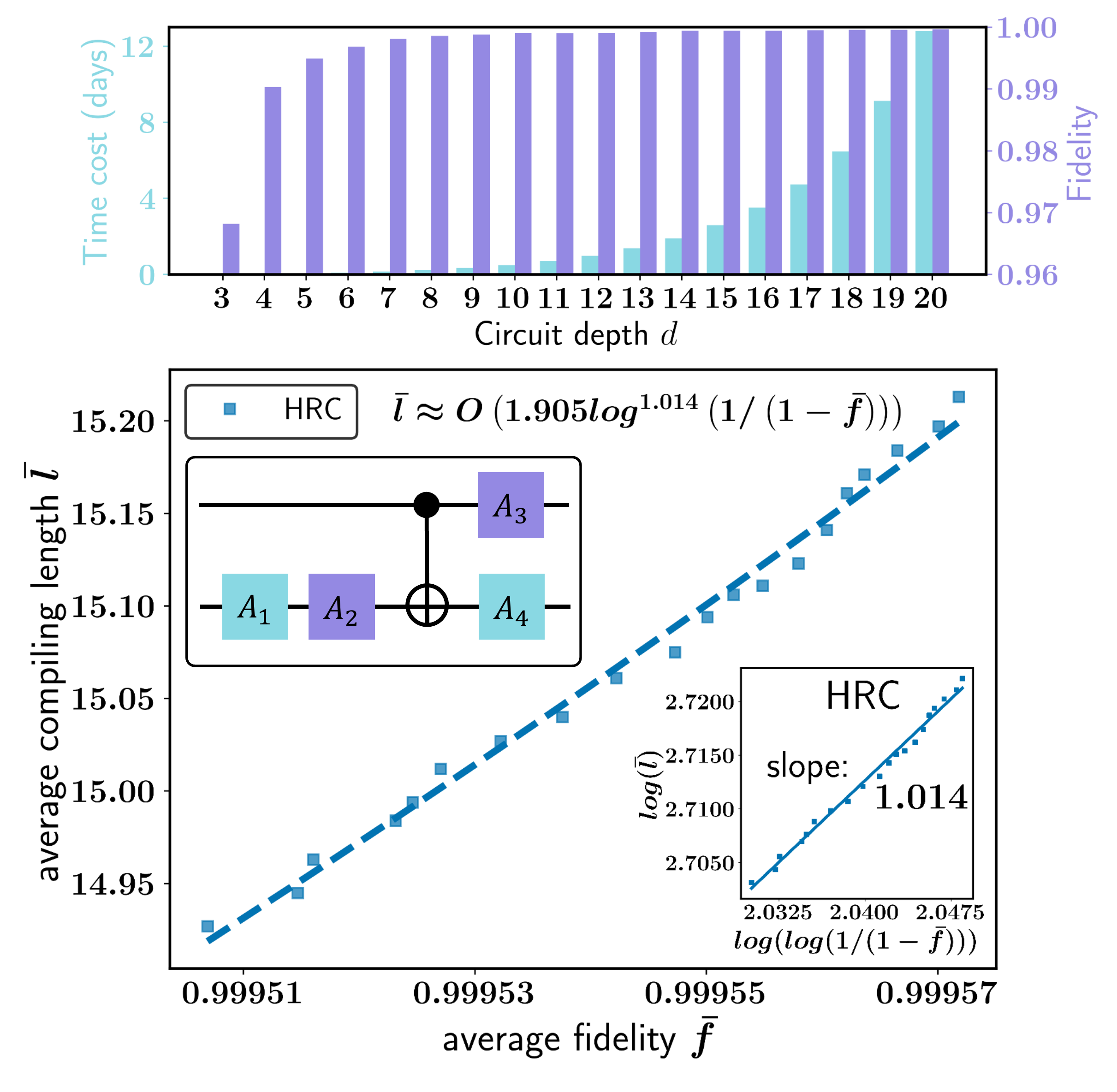}
\caption{\small{\textbf{The performance of our RL-based quantum compiler on two-qubit universal basis set.}
(a) The light-colored portion of the bar chart records the time cost of the DQN expanded by two-qubit universal basis set with gate sequence length $d$; and the dark part records the compiling fidelity $M_2$ induced by the DQN with gate sequence length $d$.
(b)Every data point is obtained by averaging the length of the approximating sequence generated by AQ* search using a validation set of $10^3$ unitary targets. We check the linearity between $\log\left(\log\left(1/\varepsilon \right)\right)$ and $\log\left(\bar{l}\right)$ in the subplot at the bottom right. We plot the output length complexity of the two-qubit universal basis set as a function of average fidelity $\bar{f}$.}
}
\label{fig:ex_2}
\end{figure}
Once the training is completed,  the proposed quantum compiler can be used to infer the gate sequence of the unseen quantum operators via the AQ* search \cite{agostinelli2021a}.  Celebrated by the ability of parallel testing all actions in  $\mathcal{A}$, the inference time of our proposal is ignorable, which ensures its scalability. To attain the optimal trajectory $\tau^{*}$ starting from any target $U$, we define an evaluation function  
\begin{eqnarray}
f(s, a; U) = G(s;U) + Q(s, a;\theta)
\label{eq:aq_search}
\end{eqnarray}
where $G(s;U)$ is the actual cumulative reward from the target unitary $U$ to the state $s$, and $Q(s, a;\theta)$ refers to DQN approximating the maximum cumulative reward from the state $s$ to $I$ taking action $a$. The essence of $f(s, a; U)$ is to provide effective guidance for finding the shortest gate sequence approximating the target unitary $U$. Concretely, during the search, we start with a set of the intermediate states and the target unitary, i.e., $\{U\prod_{j}A_j^{\dagger}, U\}$. Then, we iteratively pick the action with the maximum reward $A=\arg\max_a  f(s, a;U)$ for each $s$ and substitute $s$ with its next state $s^{\prime}$ induced by the action $A$  according to the transition probability distribution ${P}$. Once the distance between a state in $\{s\}$ and the identity state $I$ is below a designated termination accuracy $\varepsilon$,  the desired sequence between $U$ and $I$ within the tolerance threshold is obtained. The AQ* Search takes advantage of DQNs and thus outperforms other schemes, as detailed in Appendix \ref{append:aq}.

\begin{figure}
\centering
\includegraphics[width=0.49\textwidth]{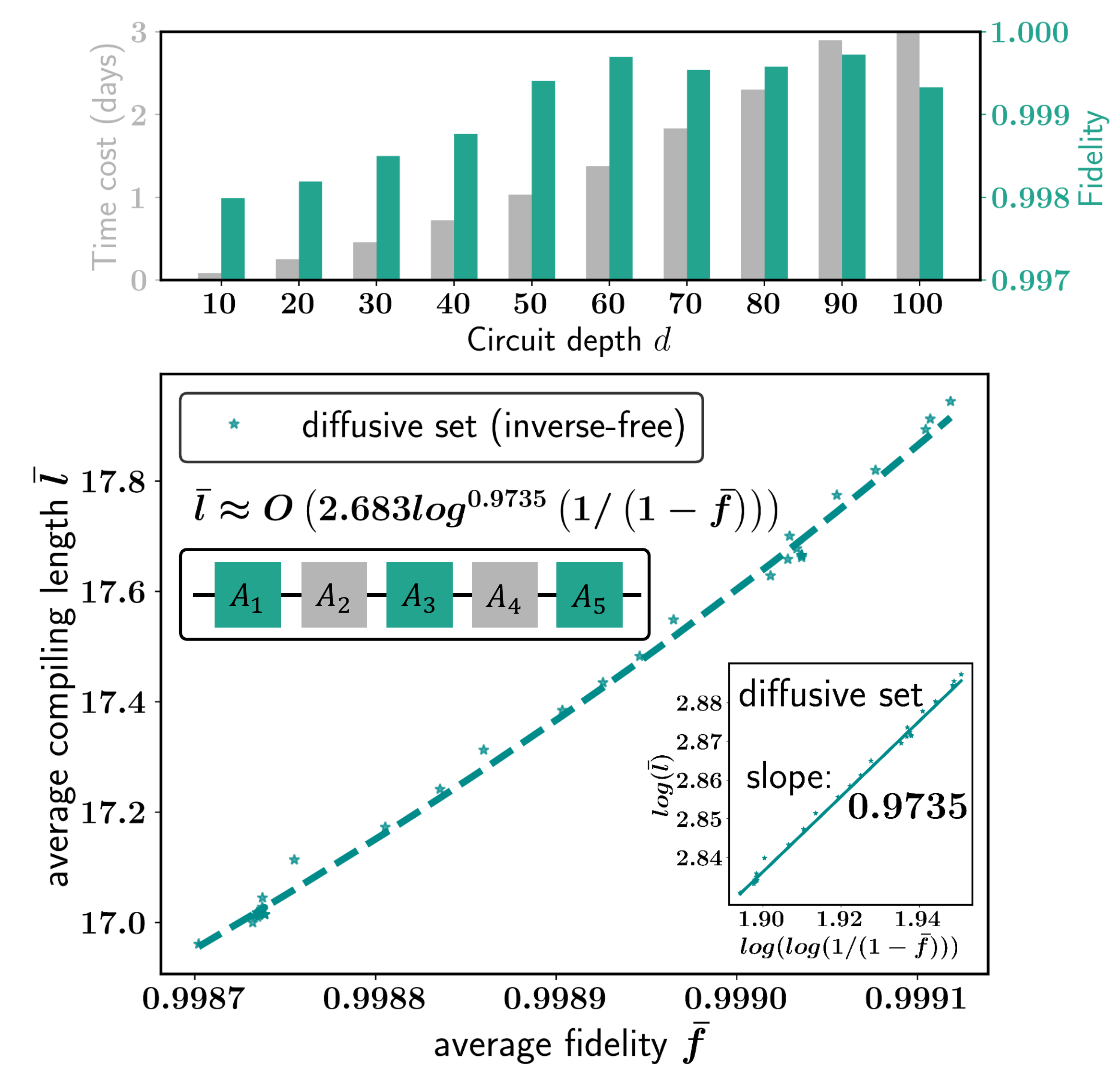}
\caption{\label{fig:ex_3} \small{\textbf{The performance of our RL-based quantum compiler on inverse-free universal basis set.}
(a) The light-colored portion of the bar chart records the time cost of the DQN expanded by the diffusive universal basis set with gate sequence length $d$; the dark part records the compiling fidelity $M_2$ induced by the DQN with gate sequence length $d$.
(b)Every data point is obtained by averaging the length of the approximating sequence generated by AQ* search using a validation set of $10^3$ unitary targets. We check the linearity between $\log\left(\log\left(1/\varepsilon \right)\right)$ and $\log\left(\bar{l}\right)$ in the subplot at the bottom right. We plot the output length complexity of the diffusive universal basis set as a function of average fidelity $\bar{f}$.}
}
\end{figure}

\section{Simulation results.} 
We conduct extensive experiments to exhibit the effectiveness of our proposal.  To examine the universality and the efficiency of multi-qubit operators, we separately apply our proposal to compile inverse-closed single-qubit and  two-qubit operators with different universal basis sets.  Furthermore, to address a long-standing problem in the inverse-free Solovay-Kitaev theorem, we apply our proposal to compile single-qubit operators under an inverse-free basis to seek a lower exponent $c$.

\textit{\underline{Inverse-Closed Universal Gates}.}
To demonstrate the versatility of our proposal, we focus on three different inverse-closed universal basis sets to make single-qubit compiling, i.e., \textit{Clifford}+\textit{T} group, Fibonacci anyons, and HRC universal basis set. In the following, we evaluate the performance of our RL-based quantum compiler by separately compiling single-qubit and multi-qubit operators using these basis sets.

The protocol setting is as follows. The employed DQN $Q(s,a;\theta)$ consists of two hidden layers, six residual blocks, and $|\mathcal{A}|$ output neurons, where $|\mathcal{A}|=5, 6, 4$ for \textit{Clifford}+\textit{T} group, HRC gates, and  Fibonacci anyons, respectively. We exploit the {Adam} \cite{kingma2017adam} optimizer to optimize DQN, and set learning rate as $\eta =10^{-3}$ without weight decay. We set the accuracy threshold in Eq.~(\ref{eqn:QCL-def}) as $\varepsilon = 10^{-3}$ and the threshold of the mean square error loss in Eq.~(\ref{eqn:bellman_loss}) as $\delta = 10^{-2}$. The gate sequence length $d$ in Fig.~\ref{fig:data_generation+dqn}(a)   varies from 3 to 40. In the inference stage, the number of test samples is set as  $10^3$. See Appendix \ref{append:architecture} for the omitted details.

To quantitively measure the efficiency of our protocol under different basis sets, we exploit two common metrics in quantum compiling, i.e., Frobenius norm 
\begin{eqnarray}
M_1(U_n, U)&=&|| {U_n}^{\dagger}U - I ||_F \nonumber \\
~&=& \sqrt{\operatorname{Tr}\left((U_n^{\dagger}U-I)(U_n^{\dagger}U-I)^{\dagger}\right)} \nonumber \\
~&=& \sqrt{\operatorname{Tr}(2I-U^{\dagger}U_n-U_n^{\dagger}U)},
\label{eq:f_norm}
\end{eqnarray}
and fidelity
\begin{eqnarray}
M_2(U_n, U)=\int \langle\psi|{U_n}^{\dagger} U| \psi\rangle 
\langle\psi|U^{\dagger} {U_n}| \psi\rangle d \psi,
\label{eq:fidelity}
\end{eqnarray}
with Haar measure $\int d \psi=1$. 
The metric $M_1$ is measures the distance between two matrices and its derivatives have been used in many quantum subfields \cite{Zhang_2020, watrous2009semidefinite}. The metric $M_2$ concerns the fidelity between the complied and target unitary \cite{moro2021quantum}.

The simulation results of compiling single-qubit operators are shown in Fig.~\ref{fig:ex_1}. In particular, Figs.~\ref{fig:ex_1}(a)-(c) exhibit  the simulation results under different basis sets. Particularly, the average fidelity of the HRC universal basis set, \textit{Clifford}+\textit{T} group, and the \textit{Clifford}+\textit{T} group is $\overbar{M_2}=99.999\%$ when $d=15$, $\overbar{M_2}=99.996\%$ when $d=40$, and $\overbar{M_2}=99.88\%$ when $d=40$, respectively. According to the two metrics and the learning-time consumption,  the HRC set  outperforms the rest two universal basis sets. The inferior performance of the \textit{Clifford}+\textit{T} group is prohibited by its sparsity. Fig.~\ref{fig:ex_1}(b) depicts the compiled gate sequence during the training process. That is, with increasing the gate sequence length $d$, the compiling accuracy is constantly enhanced.

We further utilize the simulation results to infer the sequence length complexity of our RL-based quantum compiler via extrapolation. With setting $\varepsilon = 1-\overbar{M_2}$, the sequence length complexity of our RL-based quantum compiler scales with $O\left(0.89log^{1.077}(1/\varepsilon)\right)$ for the HRC gates, $O\left(0.737log^{1.52}(1/\varepsilon)\right)$ for the Fibonacci anyons, and $O\left(2.196log^{1.25}(1/\varepsilon)\right)$ for the \textit{Clifford}+\textit{T} group, respectively. An illustration is    demonstrated in Fig.~\ref{fig:ex_1}(d) and a comparison with other quantum compilers is summarized in Table.~\ref{tab:table1}. Specifically,  our RL-based compiler outperforms prior RL-based compilers and is near-optimal, guaranteed by the inverse-closed Solovay-Kitaev Theorem. All of these observations validate the effectiveness of our proposal. More simulation results are provided in Appendix \ref{appendix:f_norm_single_qubit}.

The simulation results of compiling two-qubit operators are shown in Fig.~\ref{fig:ex_2}. Here we only  exploit the HRC universal basis set and the CNOT gate. Compared to the single-qubit case, most of the hyperparameters settings keep unchanged during training, except for the structure of DQN, which is detailed in Appendix \ref{append:architecture}. After training, the sequence length complexity of our RL-based compiler  scales with $O\left(1.905log^{1.014}\left(1/\left(1-\bar{f}\right)\right)\right)$ with fidelity above $0.9995$, as demonstrated in Fig.~\ref{fig:ex_2}(b). This empirically indicates that our RL-based compiler approaches to the optimal compiler under the HRC universal basis set in the two-qubit case. These results verify the efficiency of the HRC universal basis set in compiling multi-qubit operators. Refer to Appendix~\ref{appendix:sup_two_qubit} for the details and examples of our decomposition.

\textit{\underline{Inverse-Free Universal Gates}.}
We next apply our proposal to complete  single-qubit operator compilation using an inverse-free diffusive basis set. The exploited basis set \cite{zhiyenbayev2018quantum} is composed of the gates $\mathcal{A}=\{\widehat{A}, \widehat{B}\}$, where  $\widehat{A}=\widehat{H} \cdot \widehat{F}$, $\widehat{B}=\widehat{T} \cdot \widehat{F}$,  $\widehat{H}$ is the Hadamard gate, $\widehat{T}$ the $T$-gate, and $\widehat{F}$ a randomly generated unitary matrix detailed in Appendix \ref{ap:basis_set}. During the training stage, the setup of our protocol is identical to those introduced in the inver-closed scenario. The simulation results are  demonstrated in Fig.~\ref{fig:ex_3}. The sequence length complexity of our RL-based compiler scales with $O\left(2.683log^{0.974}\left(1/\left(1-\bar{f}\right)\right)\right)$ and the fidelity $\overbar{M_2}$ is above $0.9987$.  
 
The achieved results also address a long-standing problem in the inverse-free Solovay-Kitaev theorem, which is designing an optimal compiling algorithm with the lowest exponent $c$. Recall that under the same diffusive set presented above, the most advanced inverse-free quantum compiling algorithm generates a sequence with length complexity $c=\log 3 / \log 2$ \cite{zhiyenbayev2018quantum}. According to the achieved simulation results, our protocol allows a lower exponent $c$ than this deterministic solution. A recent study proposed an inverse-free Solovay-Kitaev theorem \cite{bouland2021efficient}. That is, in the single-qubit situation,  the sequence length for the inverse-free universal basis sets scales with $O(log^c(1/\varepsilon))$ with $c=8.62$. In conjunction with the achieved results and the conclusion of \cite{bouland2021efficient}, our proposal provides certain empirical evidence for the existence of a more efficient inverse-free Solovay-Kitaev theorem. We defer the discussion under spectral norm \cite{bouland2021efficient} to Appendix \ref{appendix:f_norm_single_qubit}.

\section{Discussion}

We have proposed an efficient and practical quantum compiler for multi-qubit operators. Attributed to the power of deep RL models and heuristic search, our proposal is scalable and compatible with various universal basis sets and quantum platforms. For the first time, we demonstrate how to use an RL-based compiler to compile multi-qubit operators with high accuracy. In addition, under the inverse-closed setting, the achieved empirical results imply that our proposal is near-optimal in compiling single-qubit and two-qubit operators, supported by the Solovay-Kitaev theorem. Furthermore, under the inverse-free setting, our proposal outperforms the current state-of-the-art method. This provides concrete empirical evidence for pursuing a more advanced inverse-free Solovay-Kitaev theorem. Consequently, our proposal can not only accelerate the realization of large-scale quantum algorithms and quantum supremacy but also facilitates a theoretical understanding of the capabilities and limitations of quantum compiling.  

Although our main focus in this study is utilizing reinforcement learning to design efficient quantum compilers, the concept behind our proposal can be generalized to other crucial quantum learning tasks from pulse and circuits to the logical level. In particular, at the pulse level,  RL techniques can be used to solve quantum control problems \cite{Magann2021Pulse,bukov2018reinforcement,niu2018universal}; at the circuit level, RL methods have the potential to accelerate Hamiltonian simulation \cite{bolens2021reinforcement}; at the logical level, RL approaches contribute to executing complicated circuits on near-term quantum machines by reducing the number of gates \cite{fosel2021quantum,du2020quantum,kuo2021quantum}. For a specified quantum learning task listed above, how to leverage the prior knowledge to design a powerful RL model to attain a better performance deserves to be further explored.

%

\providecommand{\noopsort}[1]{}\providecommand{\singleletter}[1]{#1}%

\clearpage
\newpage
\newpage
\appendix 
\onecolumngrid

The organization of the appendix is as follows.  In Appendix \ref{appendix:QC_in_RL}, we briefly introduce the essential background of   \textit{Markov Decision Process} (MDP), reformulate the problem of quantum compiling in the framework of MDP, and review other RL-based compilers from the perspective of MDP.  In Appendix \ref{append:comp-value-based}, we explain how our proposal differs from other RL-based compilers from a theoretical perspective. For the single-qubit case, we detail the inverse-closed and inverse-free universal basis sets in Appendix \ref{ap:basis_set} and demonstrate more simulation results   in Appendix \ref{appendix:f_norm_single_qubit}.  For the two-qubit case, we detail the basis set and exhibit more simulation results in Appendix \ref{appendix:sup_two_qubit}.  
The detailed architecture of the neural networks used in our proposal is shown in Appendix \ref{append:architecture}. Last, in Appendix \ref{append:aq}, the implementation of AQ* search  is presented.

\section{\label{appendix:QC_in_RL}Quantum compiling with reinforcement learning}

\subsection{\label{appendix:MDP}Markov Decision Process}
In a standard reinforcement learning (RL) system \cite{kaelbling1996reinforcement},  the agent interacts with the environment through the try-and-error cycle to maximize the cumulative reward. As a result, to describe the dynamic of the environment, the discounted Markov Decision Process (MDP), as a reasonable learning protocol, is broadly used in   RL literature \cite{sutton2018reinforcement}.

Formally, a discounted MDP is termed as a 5-tuple $(\mathcal{S}, \mathcal{A}, \mathcal{P}, \mathcal{R}, \gamma)$, where $\mathcal{S}$ refers to a measurable state space, $\mathcal{A}$ denotes a measurable action space, $\mathcal{P}: \mathcal{S} \times \mathcal{A} \rightarrow \mathcal{M}(\mathcal{S})$ indicates the Markovian transition probability distribution from one state to the next taking a specified action $a\in\mathcal{A}$, $\mathcal{R}: \mathcal{S} \times \mathcal{A} \rightarrow \mathcal{M}(\mathbb{R})$ is the immediate reward distribution associated with this transition, and the discount factor $0 \leq \gamma<1$ is introduced to guarantee the convergence of the accumulated reward ($\gamma$ can be $1$ for the finite-horizon problems). 

We now introduce the basic mechanism of a given MDP. First, an agent starts at a state $s_{0}$. Then, at each iteration $t$, the agent takes an action $a_{t}\in \mathcal{A}$ guided by a policy $\pi(\cdot \mid s_{t})$. Next, the agent observes the next state $s_{t+1}$ obeying the transition probability distribution $P(\cdot \mid s_t, a_t)$ and obtains the immediate reward $r_{t}=r(\cdot \mid s_t, a_t)$. This completes one iteration. After repeating $T$ iterations, the historical interaction path $\tau=(s_0, a_0, r_1, s_1, a_1, r_2, \cdots, s_{T-1}, a_{T-1}, r_{T-1})$ is named as a \textit{trajectory}. To maximize the cumulative reward $\sum_{j=1}^{\infty} \gamma^j r_j$, the agent learns the optimal policy $\pi^{*}$ in a try-and-error manner from the experienced  trajectories $\{\tau\}$ to guide its action.

\subsection{\label{appendix:QC_in_MDP}Quantum Compiling in the framework of MDP}
To compile any target $U$ within the tolerance $\varepsilon$, the agent aims to construct a unitary $U_{t}=\prod_{j=t-1}^{0} A_{j}$ from a discrete universal basis set $\mathcal{A}$ in the sense that the composition of $U_{t}$ uses the minimum number of basis gates in $\mathcal{A}$ and satisfies $d(U U_t^{\dagger}, I)<\varepsilon$.  This problem of discrete optimization can be cast to MDP  \cite{alam2019quantum}. In particular,  the action space $\mathcal{A}$ can be defined as the hardware-specific universal operators;  the state space $\mathcal{S}$ is referred to as the unitary space $\{S_t=U\cdot U_t^{\dagger}\}$, which encodes all of information needed by the agent to execute  compiling;  the transition probability distribution $\mathcal{P}$ is referred to as a deterministic distribution  ${P}(S_{t+1}=UU_{t+1}^{\dagger} \mid S_t, A_t)=1$, where $U_{t+1}=A_{t} \cdot U_t=\prod_{j=t}^{0} A_{j}$. Meanwhile, the Markov property of this transition can be formulated as  ${P}\left[S_{t+1} \mid S_{t}, A_{t}\right]={P}\left[S_{t+1} \mid S_{1}, A_{1}, S_{2}, A_{2}, \dots, S_{t}, A_{t}\right]$, which means that the distribution of the next state  only depends on the present state-action pair and is independent of the previous state-action pairs. Without loss of generality, we treat the cost of different operators $A_j$ chosen from the action space $\mathcal{A}$ to be equal and thus obtain a general reward function  
\begin{eqnarray}
r\left(S_{t-1}, A_{t-1}\right)= \begin{cases} 0 & \text { if } d\left(S_{t}, {I}\right)<\varepsilon \\ -1 & \text { otherwise}\end{cases},
\label{eqn:reward}
\end{eqnarray}
where $S_{t}$ is the next state following the deterministic transition probability distribution $\mathcal{P}$.  The interaction between the agent and the environment terminates at $r(S_{t-1}, A_{t-1})=0$, which means $d\left(S_{t}, {I}\right)<\varepsilon$.

Following the above explanations, the aim of the optimal quantum compiler is finding the optimal trajectory $\tau^*=(S_0, A_0, R_1, S_1, \cdots, S_{t-1}, A_{t-1}, R_{t}, S_{t})$ with the maximum cumulative reward $\sum_{j=1}^{t} \gamma^j R_j$ and $\gamma=1$.  The Solovay-Kitaev theorem guarantees the existence of the composite unitary $U_{t}=\prod_{j=t-1}^{0} A_j$ satisfying $d(S_t, I)<\varepsilon$ for any tolerance $\varepsilon$. For the optimal trajectory $\tau^{*}$, the state starts from the target unitary $S_0=U$ and ends at $S_t=UU_t^{\dagger}$ satisfying $d(S_t, I)<\varepsilon$. This implies  $S_t \approx I$ and $U\approx U_t$.  The maximum cumulative reward amounts that the agent accomplishes the quantum compiling with the minimum number of  operators below the precision threshold $\varepsilon$ and the sequence  length complexity arrives at the lower bound $O(log(1/\varepsilon))$.

We note that finding the optimal trajectory of the above MDP is extremely challenged, due to the exponential state space and sparse reward \cite{riedmiller2018learning}. Concretely, for a single-qubit system whose universal basis set contains $6$ basis gates, when $t=30$, the state space scales with $6^{30}\approx 10^{23}$;  for a two-qubit system whose universal basis set contains $14$ basis gates, when $t=30$, the state space scales with $14^{30} \approx 10^{34}$.  Due to this exponentially large space, the condition $d\left(S_{t+1}, {I}\right)<\varepsilon$ in Eq.~\eqref{eqn:reward} is hard to satisfy, which incurs $r=-1$ along the trajectories without useful information. Therefore, an untrained agent would may fail to see a positive reward signal and  unlikely outputs a demanded gate sequence  through the random trial and error strategy. To avoid the issue of the sparse reward, our proposal adopts the value-based RL methods to find the optimal policy $\pi^{*}$. Namely,  by approximating the optimal action-value function \[Q^{*}(s,a) = \max _{\pi} \mathbb{E}\left[\sum_{j=0}^{\infty} \gamma^jR_j \mid s_{0}=s, a_{0}=a, \pi\right],\] which estimates the maximum cumulative reward starting from state $s$ and taking action $a$, the agent can gain   non-zero rewards.

\begin{figure*}
\centering
\includegraphics[width=17cm]{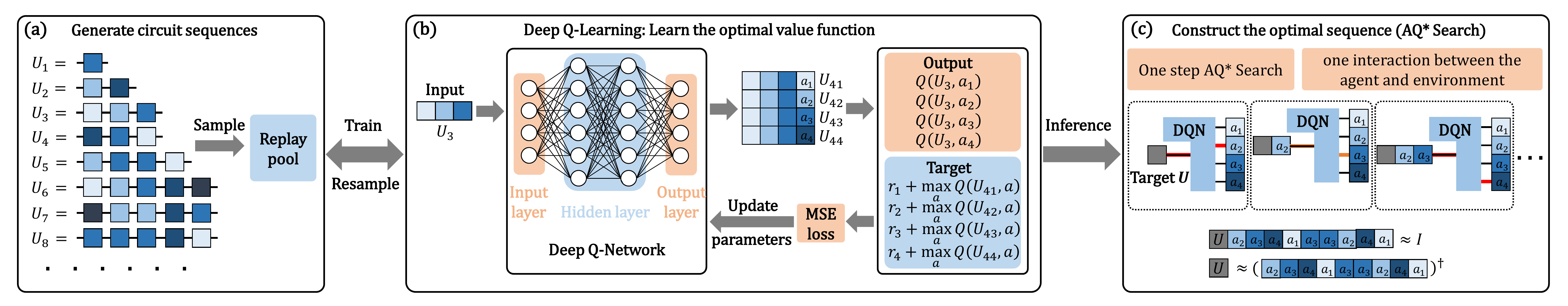}
\caption{\label{fig:scheme} \small{\textbf{An overview of the deep Q-learning quantum compiling algorithm with AQ* search.} (a) Data generation. The behavior distribution is constructed from the sample trajectories. (b) Deep Q-learning. In the training stage, the deep Q-network is introduced to approximate the optimal value function and optimal strategy. (c) Construct the optimal sequence. In this inference stage, we implement the AQ* search guided by the trained deep Q-network to interact with the environment.}}
\end{figure*}

\subsection{\label{appendix:review_DRL}A brief review of deep reinforcement learning oriented quantum compiling}
Here we recap the essential background of deep reinforcement learning for self-consistent. Refer to Refs.~\cite{sutton2018reinforcement, bertsekas2019reinforcement} for the comprehensive review.

From the perspective of the optimization target, the RL methods can be divided into the value-based methods \cite{arulkumaran2017deep}, which engage the Bellman equation to approximate the optimal value function, and policy-based methods, which exploit the policy gradient \cite{sutton2018reinforcement, nachum2017bridging} to direct approximate the optimal policy. According to the setup of the transition probability distribution $\mathcal{P}$,   RL methods can be divided into the model-free methods \cite{glascher2010states}, which learn value function or policy through trial and error interaction with a black-box environment, and model-based methods \cite{berkenkamp2017safe}, which learn from high-reward-prone experience by leveraging the transition probability $\mathcal{P}$.

To alleviate the issues of the exponentially large state space and the sparse reward in quantum compiling, initial studies \cite{moro2021quantum, gong2021weighted} have explored the potential of value-based and policy-based RL algorithms in the model-free setting. Namely, when compiling a single-qubit operator, prior methods assisted by certain tricks, e.g., hindsight experience replay \cite{andrychowicz2018hindsight}, attain good performance. Nevertheless, these methods encounter the phenomenon of sparse reward, which forbids their scalability to the multi-qubit situation. Besides,  numerical results indicate that the generated compiling sequences of these proposals are not optimal.

Beyond the model-free methods, Ref.~\cite{Zhang_2020} exploits model-based methods, which generate many sample trajectories with positive reward signals according to the transition probability $\mathcal{P}$ and eliminates the sparse reward phenomenon. Although the proposed method can generate `near-optimal' gate sequences, it can only be applied to the single-qubit case due to the huge computational overhead. Ref.~\cite{moro2021quantum} exploits   $d(U_n, U)$ as the reward function. However, this approach cannot warrant the optimality guarantee in solving large-scale quantum problems.

Our proposal differs from previous RL-based quantum compilers. As shown in Fig.~\ref{fig:scheme}, during the data generation before training, our proposal captures the prior information of the quantum compiling problems and customizes the behavior distribution, which avoids the dilemma of sparse reward. In addition, during the training stage, our proposal exploits a deep Q-network, which can capture more useful information to approximate the optimal strategy compared with Ref.~\cite{Zhang_2020}. Moreover, the employed AQ* search takes full advantage of the learned deep Q-network and ensures the superior performance of our proposal over other inference strategies.   
  
\section{Comparison with other value-based RL methods}\label{append:comp-value-based}
The value-based RL algorithms focus on detecting optimal policy $\pi^{*}$ by  approximating  the optimal value function.  In the context of RL, there are two kinds of value function, i.e., the state-value function $V^{\pi}(s)=\mathbb{E}\left[\sum_{j=0}^{\infty} \gamma^jR_j \mid s_{0}=s, \pi\right]$ that outputs the cumulative reward starting from state $s$ under the policy $\pi$, and action-value function $Q^{\pi}(s, a)=\mathbb{E}\left[\sum_{j=0}^{\infty} \gamma^jR_j \mid s_{0}=s, a_{0}=a, \pi\right]$ that outputs the cumulative reward starting from the state $s$ and an action $a$ under the policy $\pi$. Bellman optimality equation for the action-value function $Q(s,a)$  in Eq.~(\ref{eq:bellman_optimal})  indicates
\begin{eqnarray}
V^{*}(s)=\max _{a} \mathbb{E}_{s^{\prime} \sim P(\cdot \mid s, a)}\left[r+\gamma  V^{*}\left(s^{\prime}\right) \middle| s\right],
\label{eq:bellman_optimal_state}
\end{eqnarray}
where $(s, a, P, r, \gamma)$ corresponds to the  state, action, transition probability distribution, reward, discount factor, respectively, and $s^{\prime}$ is the next state obeying the deterministic distribution $P(\cdot \mid s, a)$. Using this Bellman equation as an iterative update, i.e., $V^{(t)}(s)=\max_{a} \mathbb{E}\left[r+\gamma V^{(t-1)}\left(s^{\prime}\right) \middle| s \right]$, the state-value function $V^{(t)}$ also converges to the optimal state-value function obeying Eq.~(\ref{eq:bellman_optimal_state}), i.e.,  $V^{(t)} \rightarrow V^{*}$ as $t \rightarrow \infty$.

Different from Ref.~\cite{Zhang_2020} that approximates the optimal state-value function $V^{*}(s)$, we employ the action-value function $Q^{*}(s, a)$, which has advantages in both the training phase and the inference phase.
The state-value function $V(s)$ can be represented as a deep neural network, which outputs a scalar approximating the value of the input state $s$.
Meanwhile, the action-value function $Q(s,a)$ can also represented as the deep Q-network, which outputs a vector approximating the value of every next state $s^{\prime}$. While the number of parameters of the deep Q-network grows linearly with the size of the action space, the number of the forward passes needed to compute the loss function stays constant for each update. Moreover, Ref.~\cite{agostinelli2021a} has indicated that in a large action space $\mathcal{A}$, the training time for deep Q-network can be up to $100$ more times faster than the state-value function with the similar layout. Besides, in the inference phase, each query of the action-value function is equivalent to querying the state-value function with  $\left|\mathcal{A}\right|$ times. This linear reduction in the inference time is beneficial to generalize our algorithm to the multi-qubit scenario.

\section{Universal basis sets}\label{ap:basis_set}
\textbf{Fibonacci anyons basis sets.} Fibonacci anyons are quasiparticle excitations of topological states that obey non-Abelian braiding statistics \cite{Nayak_2008} and the simplest non-Abelian quasiparticles that enable universal topological quantum computation \cite{Kitaev_2006} by braiding alone \cite{freedman2002modular}. Their mathematical expression is 
\begin{eqnarray}
A_1=\left(\begin{array}{cc}
\eta^{-4} & 0 \\
0 & \eta^{3}
\end{array}\right),
\quad
A_2=\left(\begin{array}{cc}
-\phi^{-1} \eta^{-1} & \phi^{-\frac{1}{2}} \eta^{-3} \\
\phi^{-\frac{1}{2}} \eta^{-3} & -\phi^{-1}
\end{array}\right),
\label{eq:eight}
\end{eqnarray}
where $\eta = e^{i\pi/5}$ and $\phi = \frac{\sqrt{5}+1}{2}$.

\textbf{HRC basis set.}  The  HRC universal basis set proposed in \cite{Harrow_2002} takes the form
\begin{eqnarray}
&B_1 = \frac{1}{\sqrt{5}} \left(\begin{array}{cc}
1 & 2i \\
2i & 1
\end{array}\right),
\quad
B_2 = \frac{1}{\sqrt{5}} \left(\begin{array}{cc}
1 & 2 \\
-2 & 1
\end{array}\right), 
\quad
B_3 = \frac{1}{\sqrt{5}} \left(\begin{array}{cc}
1+2i & 0 \\
0 & 1-2i
\end{array}\right).
\label{eq:nine}
\end{eqnarray}

\textbf{\textit{Clifford}+\textit{T} basis set.} The $n$-qubit \textit{Clifford} group  is generated by the Hadamard gate $H$, the phase gate $S$, the controlled-not gate\cite{Giles_2013}. One can obtain a universal basis set by adding the \textit{non-Clifford} operator $T$  into \textit{Clifford} group. The mathematical expression of the basis gates is  
\begin{eqnarray}
&H = \frac{1}{\sqrt{2}} \left(\begin{array}{cc}
1 & 1 \\
1 & -1
\end{array}\right),
\quad
S = \left(\begin{array}{cc}
1 & 0 \\
0 & i
\end{array}\right), 
\quad
T = \left(\begin{array}{cc}
1 & 0 \\
0 & e^{i\pi/4}
\end{array}\right).
\label{eq:ten}
\end{eqnarray}
In the numerical simulations, considering that the training difficulty is exacerbated by the sparsity of $T$ and $S$, we replace $S$ by $H\cdot S$ in training DQN.

\textbf{Inverse-free basis set.} A limitation of the standard Solovay-Kitaev theorem is that it requires the gate set to be inverse-closed.  An alternative would be inverse-free gate sets, which are are diffusive enough such that the sequences of moderate length cover the space of unitary matrices in a uniform way \cite{zhiyenbayev2018quantum}.  To verify our proposal in the inverse-free setting, we exploit a diffusive set $\mathcal{M}$ composed of two gates $\{\widehat{A}, \widehat{B}\}$. More precisely, $\widehat{A}=\widehat{H} \cdot \widehat{F}$ and $\widehat{B}=\widehat{T} \cdot \widehat{F}$ where $\widehat{H}$ is the Hadamard gate, $\widehat{T}$ the $T$-gate and $\widehat{F}$ a randomly generated unitary matrix
\begin{eqnarray}
\widehat{F}=\left(\begin{array}{cc}
-0.40194-\mathrm{i} 0.43507 & -0.36803-\mathrm{i} 0.71674 \\
0.36803-\mathrm{i} 0.71674 & -0.40194+\mathrm{i} 0.43507
\end{array}\right).
\end{eqnarray}

 \begin{figure}
\centering
\includegraphics[width=18.0cm]{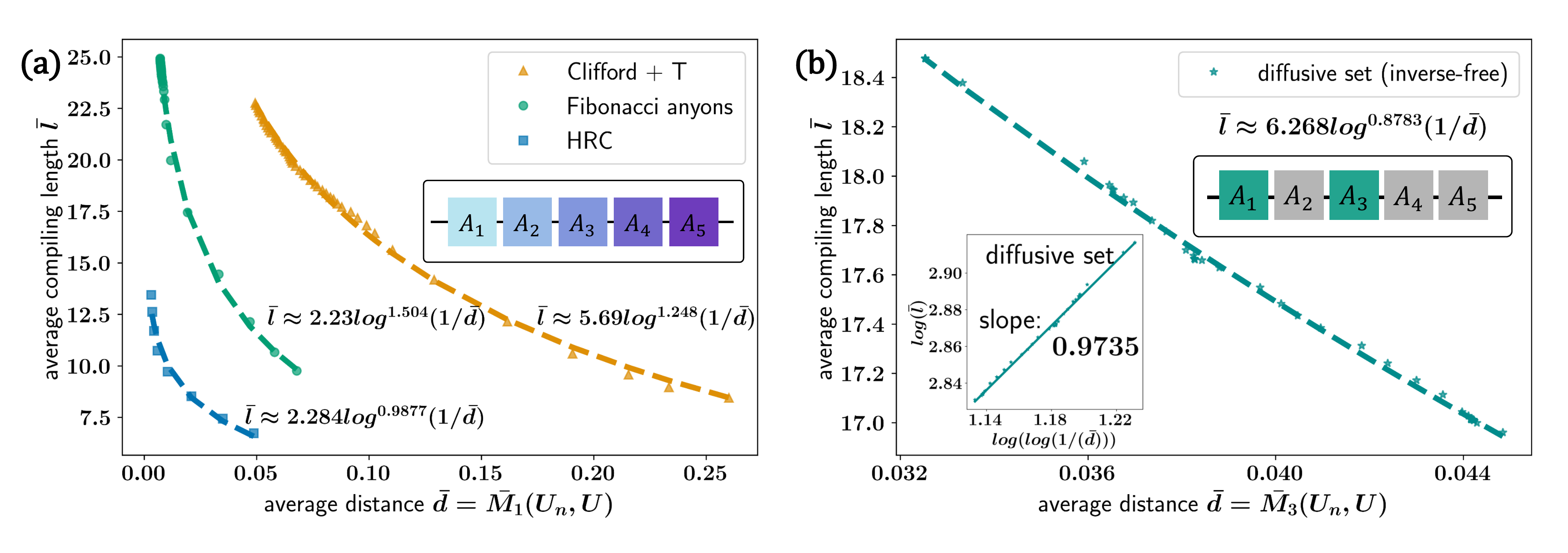}
\caption{\label{fig:f_norm}\small{\textbf{Comparison between different single-qubit universal basis sets under the metric of (a) F-norm and (b) spectral norm.} 
}}
\end{figure}

\section{\label{appendix:f_norm_single_qubit} More simulation results for the single-qubit operator compiling}
In the main text, we only discuss the length complexity of the generated gate sequences as a function of fidelity $M_2(U_n, U)=\int
\langle\psi|{U_n}^{\dagger} U| \psi\rangle 
\langle\psi|U^{\dagger} {U_n}| \psi\rangle d \psi$. Notably, some studies employ the F-norm \cite{Zhang_2020} or spectral norm \cite{bouland2021efficient} as the metric to measure the distance between $U_n$ and $U$. To this end, here we replace $M_2(U_n, U)$ bu the F-norm $M_1(U_n, U)=|| {U_n}^{\dagger}U - I ||_F$ and the spectral norm $M_3(U_n, U)=\sup _{\langle\psi \mid \psi\rangle=1} \|(U_n-U)|\psi\rangle \|_{2}$ to evaluate the performance of our proposal. 

The structure of the DQN  and  the hyperparameters settings are identical to those introduced in the main text. We set the accuracy threshold in Eq.~(\ref{eqn:QCL-def}) as $\varepsilon = 10^{-3}$ and the threshold of the MSE loss in Eq.~(\ref{eqn:bellman_loss}) as $\delta = 10^{-2}$. The gate sequence length $d$ in Fig.~\ref{fig:data_generation+dqn}(a)   varies from 3 to 40. Every data point is obtained by averaging the length of the compiling sequence generated by AQ* search using a validation set containing  $10^3$ unseen unitary targets. 

The achieved sequence length complexity of our proposal via extrapolation under the metric of F-norm $M_1$ is shown in Fig.~\ref{fig:f_norm}(a). With setting $\varepsilon = \overbar{M_1}$, the sequence length complexity of our RL-based quantum compiler scales with $O\left(2.284log^{0.9877}(1/\varepsilon)\right)$ for the HRC gates, $O\left(2.23log^{1.504}(1/\varepsilon)\right)$ for the Fibonacci anyons, and $O\left(5.69log^{1.248}(1/\varepsilon)\right)$ for the \textit{Clifford}+\textit{T} group, respectively. These results validate that our proposal is robust under different distance metrics $d(U_n, U)$. For the diffusive gates set, the achieved sequence length complexity of our proposal under the metric spectral norm $M_3$ is plotted in Fig.~\ref{fig:f_norm}(b). When $\varepsilon = \overbar{M_3}$, the sequence length complexity of our RL-based quantum compiler scales with $O\left(6.268log^{0.8783}(1/\varepsilon)\right)$. Our protocol allows a lower exponent $c=0.8783$ than $c=8.62$ in Ref.~\cite{bouland2021efficient} under same metric, providing concrete empirical evidence in pursing a more advanced inverse-free Solovay-Kitaev theorem.

\section{\label{appendix:sup_two_qubit}More simulation results for the two-qubit operator compiling}
 Most RL-based quantum compilers suffer from the exponentially large state space with respect to the number of qubits. As a result, they have to restrict the state space. In  \cite{Zhang_2020}, the number of the employed CNOT gates and their position in the complied unitary are fixed, which reduces the two-qubit operator compiling task into a single-qubit operator compiling task. Alternatively, Ref. \cite{moro2021quantum} fixed the compiling targets as the multiplication of the universal gates instead of randomly sampling from $SU(4)$. In contrast, our is universal. 
 
 We now demonstrate more simulation results omitted in the main text. In particular,   the adopted universal basis set is formed by HRC universal gates and the CNOT gates. Mathematically, the action space yields 
\begin{equation}
\begin{aligned}
    \mathcal{A}^{\prime}=\{& H\otimes I, H^{\dagger}\otimes I, I\otimes H, I\otimes H^{\dagger},   R\otimes I, R^{\dagger}\otimes I, I\otimes R, I\otimes R^{\dagger},  C\otimes I, C^{\dagger}\otimes I, I\otimes C, I\otimes C^{\dagger}, \\
    & \ket{0}\bra{0}\otimes I+\ket{1}\bra{1}\otimes X,   I\otimes \ket{0}\bra{0}+X\otimes \ket{1}\bra{1} \},
\end{aligned}
\end{equation}
where the dimension of the action space is $|\mathcal{A}|=14$. The gate sequence length  $d$ ranges from $3$ to $20$.  For ease of illustration, we set the target operator  as $U = e^{-i\frac{\alpha}{2}(X\bigotimes Z)}$ with $\alpha\in \{1, 13, 26, 44\}$.  

\begin{figure}
\centering
\includegraphics[width=18cm]{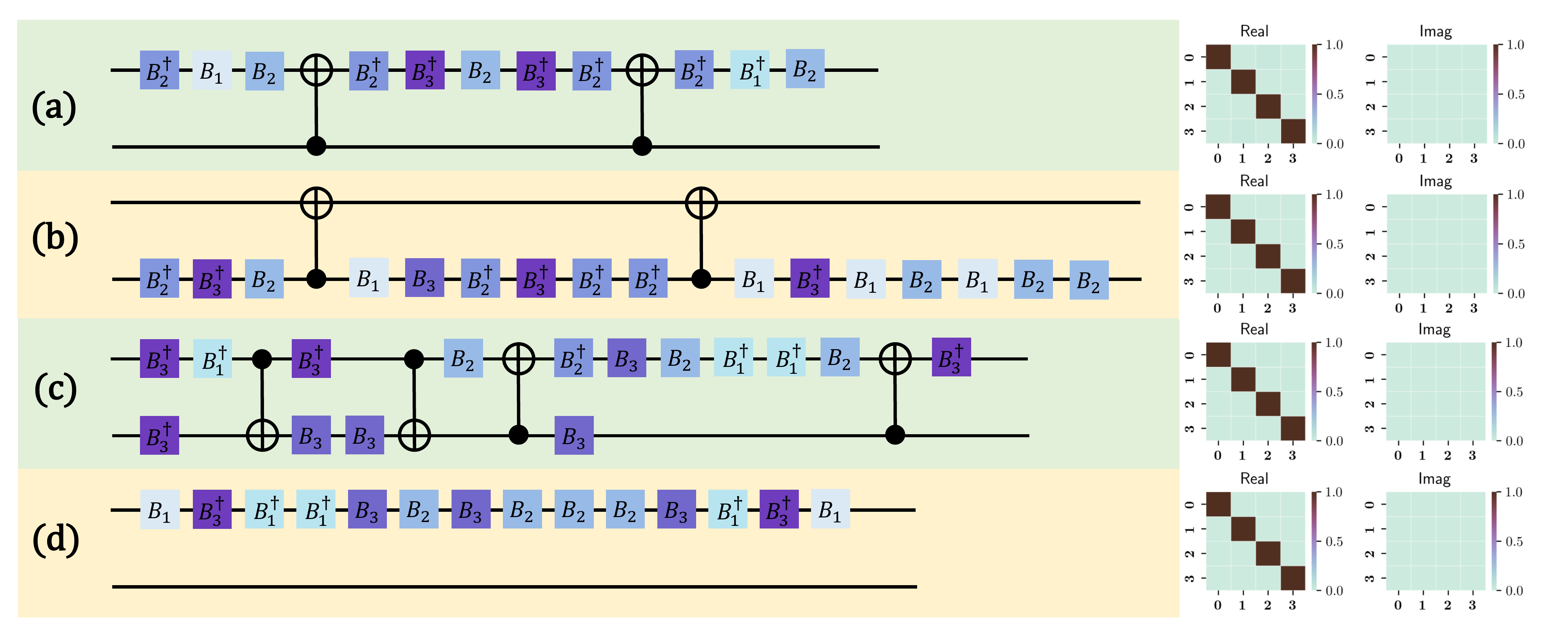}
\caption{\label{fig:appendix_ciucuit}\small{\textbf{Simulation results in compiling the two-qubit operators.} The compiling target opeator is $U = e^{-i\frac{\alpha}{2}(X\bigotimes Z)}$. Every unitary $U_n$ is generated by AQ* search guided by the learned DQN. (a) $\alpha=1$, (b) $\alpha=13$, (c)$\alpha=26$, (d)$\alpha=44$. The heatmaps on the right panel depict the real and imaginary parts of the unitary matrix $U_n^{\dagger}U$, where $U_n$ is generated from our proposal. The horizontal and vertical coordinates in the heatmaps correspond to the corresponding positions of the unitary matrix.
}}
\end{figure}

The simulation results are depicted in Fig.~\ref{fig:appendix_ciucuit}, where each row stands for the compiling result for a certain target unitary. More specifically, the circuit diagram depicts   the complied gates generated by AQ* search guided by the trained DQN. The heatmap shows the fidelity $U_n^{\dagger}U$.  In particular, Fig.~\ref{fig:appendix_ciucuit}(a) indicates that when $\alpha = 1$,  we have $M_1(U_n,U)=0.0869$, $M_2(U_n,U)=0.9996$, and
\begin{eqnarray}
U_n^{\dagger}U = 
\begin{pmatrix}
9.997e^{-1}-7.780e^{-3}i & 4.769e^{-9}-2.981e^{-9}i & 2.021e^{-2}+1.745e^{-3}i & -6.242e^{-9}+1.999e^{-8}i\\
3.332e^{-9}-2.916e^{-9}i & 9.998e^{-1}-1.280e^{-2}i & -3.650e^{-16}-7.997e^{-9}i & -1.747e^{-2}-1.745e^{-3}i\\
-2.021e^{-2}+1.745e^{-3}i & -3.577e^{-9}+2.101e^{-16}i & 9.998e^{-1}+7.780e^{-3}i & -1.943e^{-8}-2.368e^{-8}i\\
1.416e^{-8}-6.664e^{-9}i & 1.747e^{-2}-1.745e^{-3}i & -6.664e^{-9}-1.066e^{-8}i & 9.998e^{-1}+1.280e^{-2}i
\end{pmatrix}. 
\end{eqnarray}
Fig.~\ref{fig:appendix_ciucuit}(b) indicates that when $\alpha = 13$, we obtain  $M_1(U_n,U)=0.0549$, $M_2(U_n,U)=0.9998$, 
\begin{eqnarray}
U_n^{\dagger}U = 
\begin{pmatrix}
9.999e^{-1}+0.007i & 9.824e^{-3}+0.004i & 6.614e^{-5}-0.002i & -6.055e^{-3}+0.001i\\
-9.824e^{-3}+0.004i & 9.999e^{-1}+0.007i & 6.055e^{-3}+0.001i & 6.616e^{-5}+0.002i\\
6.614e^{-5}-0.002i & -6.055e^{-3}+0.001i & 9.999e^{-1}-0.007i & 9.824e^{-3}+0.004i\\
6.055e^{-3}+0.001i & 6.616e^{-5}+0.002i & -9.824e^{-3}+0.004i & 9.999e^{-1}-0.007i
\end{pmatrix}.
\end{eqnarray}
Fig.~\ref{fig:appendix_ciucuit}(c) indicates that when $\alpha = 26$,  we obtain  $M_1(U_n,U)=0.0694$, $M_2(U_n,U)=0.9997$, 
\begin{eqnarray}\setlength\arraycolsep{2pt}
U_n^{\dagger}U=\begin{pmatrix}
0.9998-0.0001i & 0 & 0.0155-0.0073i & 0\\
0 & 0.9998+0.0057i & 0 & 0.0135-0.0097i\\
-0.0155-0.0073i & 0 & 0.9999+0.0001i & 0\\
0 & -0.0135-0.0097i & 0 & 0.9998-0.0057i
\end{pmatrix}. 
\end{eqnarray}
Fig.~\ref{fig:appendix_ciucuit}(d) indicates that when $\alpha = 44$,  we obtain   $M_1(U_n,U)=0.0559$, $M_2(U_n,U)=0.9998$, 
\begin{eqnarray}\setlength\arraycolsep{2pt}
U_n^{\dagger}U=\begin{pmatrix}
0.9998-0.0047i & 0 & -0.0064-0.0183i & 0\\
0 & 0.9999-0.0048i & 0 & -0.0064-0.0006i\\
0.0064-0.0183i & 0 & 0.9998+0.0047i & 0\\
0 &  0.0064-0.0006i & 0 & 0.9999+0.0048i
\end{pmatrix}. 
\end{eqnarray}

\section{Deep Q-network architecture}\label{append:architecture}
\textbf{Deep Q-Learning.} \label{Methods: DQN_details}
To optimize the loss function in Eq.~(\ref{eqn:bellman_loss}) with a stable manner, the parameters from the previous iteration $\theta^{(t-1)}$ are fixed when optimizing the loss function $L_{t}\left(\theta^{(t)}\right)$.  The gradients of $L_{t}$  with respect to the trainable parameters yield 
\begin{eqnarray}
\nabla_{\theta^{(t)}} L_{t}\left(\theta^{(t)}\right)= && C\cdot \mathbb{E}_{s, a \sim \phi(\cdot)}\left[\left(y_{t} -Q\left(s, a ; \theta^{(t)}\right)\right)\right. \nonumber\\
&&
\left. \cdot \nabla_{\theta^{(t)}} Q\left(s, a ; \theta^{(t)}\right)\right].
\label{eq:gradient}
\end{eqnarray}
Rather than computing the full expectations, it is often computationally efficient to optimize the loss function by stochastic gradient descent. To this end,  we sample  state-action pairs from the behavior distribution $\phi(s,a)$ and collect the corresponding rewards, and finally store them into a fixed-size \textit{replay pool} as shown in Fig.~\ref{fig:data_generation+dqn}(a), which avoids the construction of the huge state space and accelerates the learning process through breaking up the correlation within $(s, a)$ pairs in a trajectory. Note that the behavior distribution $\phi(s,a)$ can be manually fabricated. Here we randomly select the action sequence $\{A_j; j=0, 1, \cdots ,d-1 \}$ with the length $d$ and construct the trajectory $\tau = \{S_0, A_0, R_1, S_1, A_1, R_2, \cdots, S_{d-1}, A_{d-1}, R_d \}$ induced by these actions. At each iteration $t$,  the samples in the replay pool are used to calculate the gradient Eq.~(\ref{eq:gradient}). If the loss is beyond a pre-fixed threshold $\delta$, we empty the replay pool and resample $(s, a)$ pairs until the loss is below $\delta$.   In this way, the distribution of the sample pairs $\phi(s, a)$ from trajectory $\tau$ approximate a uniform distribution over all $(s, a)$ pairs as $d \rightarrow \infty$. This is because  the action strings  $\{A_0, ..., A_{d-1}\}$ can cover the space of unitary uniformly by increasing the gate sequence length $d$.   Moreover, any state sequence can be represented as a trajectory without sparse reward. For example, when the target unitary is $U=HSS$ and we have a action string  $\{I, H, HS, HSS\}$ with $d=3$, the corresponding trajectory yields $\tau=\{S_0=HSS, A_0=S, R_1=-1, S_1=HS, A_1=S, R_2=-1, S_2=H, A_2=H, R_3=0, S_3=I\}$, where the constructed gate sequence $U_3=\prod_{j=2}^0 A_j=HSS$ achieves a non-negative reward $R_3=0$.

The action-value function $Q(s,a;\theta)$ is represented by a deep neural network, which consists of two hidden layers, six residual blocks, and $|\mathcal{A}|$ output neurons. The first two hidden layers are of sizes 5000 and 1000 for single-qubit cases, 6000 and 2000 for multi-qubit cases, respectively, and each residual block consists of two hidden layers with 1000 hidden neurons each, as shown in Fig.~\ref{fig:architecture}.
During the training process, as the demonstration of efficiency and practicality, we keep the time consumption of the training phase within an acceptable range: no more than one week.
This thrift generates a separation between the actual DQN we learned within a fixed training time and the best DQN our algorithm can learn, and this separation gets smaller with training time.
\begin{figure}
\centering
\includegraphics[width=9.0cm]{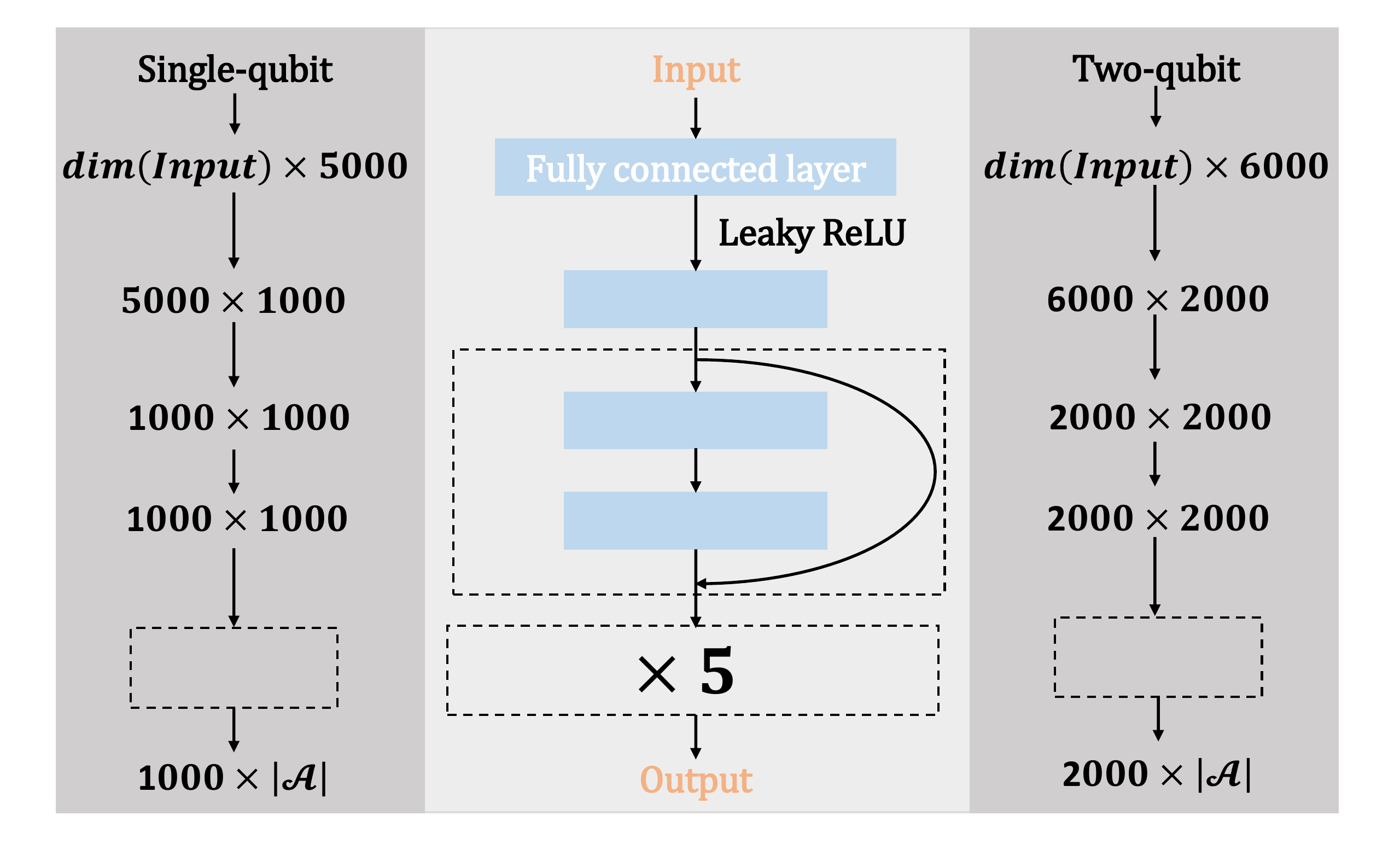}
\caption{\label{fig:architecture}\textbf{\small{The architecture of the deep Q-network for single-qubit and multi-qubit situations.}}
}
\end{figure} 

\textbf{The  gate sequence length  in inference.}
To determine the threshold of the gate sequence length $d$ in the inference stage, we continuously conduct AQ* search guided by the trained DQNs with a deeper gate sequence length. This process is completed when the change of compiling accuracy, or equivalently the change of loss function in Eq.~(\ref{eqn:bellman_loss}), becomes subtle.

\section{More details of AQ* Search}\label{append:aq}
The Pseudo code of AQ* search is summarized in Algorithm.\ref{al:aq}.  Unlike conventional approaches that simply normalize the learned deep Q-network to derive the policy,  AQ* search  facilitates the generation of the optimal trajectory by combining the DQN with search methods, which is widely used in high-dimensional complex control problems (e.g., Go \cite{silver2016mastering}, Rubik’s cube \cite{agostinelli2019solving}). To exhibit the power of AQ* search, in the following, we compare the performance of AQ* search with two typical conventional policies. The first one is a random policy, $\pi_{g}(a\mid s)=\underset{a}{\operatorname{argmax}} Q(s, a;\theta)$, which constantly performs the action that is believed to yield the highest expected reward. The second one is a Boltzmann policy, i.e., $\pi_{b}(a\mid s)=\frac{e^{-Q(s, a;\theta) /(k T)}}{\sum_{j=1}^{\left|\mathcal{A}\right|} e^{-Q(s, a_{j};\theta) /(k T)}}$, where $k$ is the Boltzmann constant, $T$ is the temperature. 

We compare the efficiency of different inference strategies under single-qubit \textit{Clifford}+\textit{T} universal basis set. The hyper-parameters settings are as follows. All the inference algorithms are guided by the learned DQN with gate sequence length $d=40$. The the random strategy is realized by using the Boltzmann distribution with setting $T=1/k$. We sample $100$ single-qubit unseen unitaries and employ the above three inference algorithms to build the compiling strings.

The simulation results in the measure of fidelity are illustrated in  Fig.~\ref{fig:interaction}. The average fidelity scales with $0.9955$ for the AQ* search, $0.9784$ for the Boltzmann distribution strategy, $0.9708$ for the random strategy. From the perspective of robustness, AQ* search has less variance and thus is more stable. These results validate that AQ* search is a more powerful inference strategy compared with conventional inference strategies. 

\begin{figure}
\centering
\includegraphics[width=8.0cm]{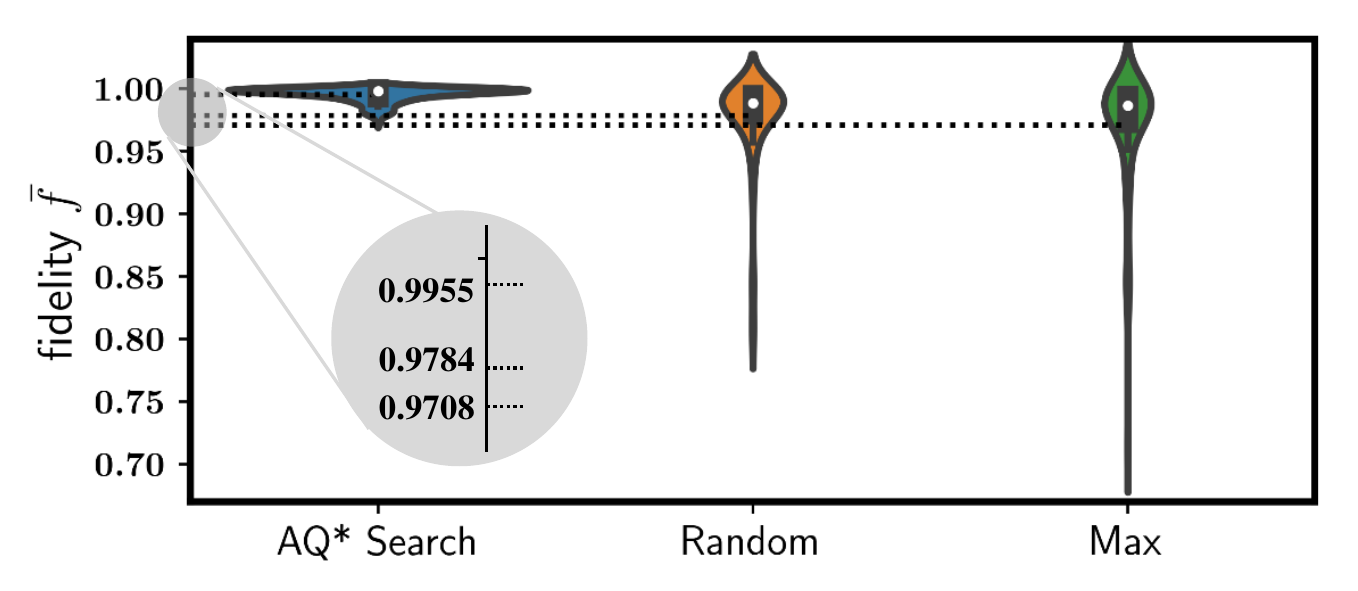}
\caption{\label{fig:interaction}\small{\textbf{Comparison the performance of AQ* search with other methods}. }
}
\end{figure}

\begin{algorithm}[H]
\caption{$\mathrm{AQ}^{*}$ Search} 
\label{al:aq}
\begin{algorithmic}[1]
\Require starting state $s_0$, deep Q-network $Q({\theta})$
\State OPEN $\longleftarrow$ priority queue
\State CLOSED $\longleftarrow$ dictionary that maps nodes to path costs
\State $n_0$ = NODE$(s=s_0, g=0)$ \Comment{$n_0 \in$CLOSED}
\State $a_0$ = $\arg\max _{a^{\prime} \in \mathcal{A}} Q(s_0, a^{\prime}; \theta)$
\State $v_0$ = $Q(s_0, a_0;\theta)$
\State Push $(s_0, a_0)$ to OPEN with reward $0+v_0$ \Comment{$(s_0, a_0) \in$OPEN}
\While{not IS$\_$EMPTY(OPEN)}
    \State $(s, a)$ = POP(OPEN)
    \State $s^{\prime} = A(s, a)$
    \If{IS$\_$GOAL($s^{\prime}$)}
        \State \Return PATH$\_$TO$\_$GOAL($s, a$)
    \EndIf
    \State $g^{\prime}=n.g+r^a(s, s^{\prime})$
    \State $n^{\prime}=$NODE$(s=s^{\prime}, g=g^{\prime})$
    \If{$n^{\prime}$ not in CLOSED or $g^{\prime}<$CLOSED$[n^{\prime}]$}
        \State CLOSED$[n^{\prime}]=g^{\prime}$
        \For{$a^{\prime} \in \mathcal{A}$}
            \State $q^{\prime}=V_{\theta}(s^{\prime}, a^{\prime})$
            \State $v^{\prime}=g^{\prime}+q^{\prime}$
            \State Push $(s^{\prime}, a^{\prime})$ to OPEN with reward $v^{\prime}$
        \EndFor
    \EndIf
\EndWhile
\State \Return Failure
\end{algorithmic} 
\end{algorithm}

\end{document}